\documentclass[a4paper]{article}

\usepackage{INTERSPEECH2022}
\usepackage{cite}
\usepackage{subcaption}
\usepackage{comment}
\usepackage{url}

\newcommand{\spcell}[2][c]{%
  \begin{tabular}[#1]{@{}c@{}}#2\end{tabular}}
  
\newcommand{\bW}{\ensuremath{\mathbf{W}}}
\newcommand{\bL}{\ensuremath{\mathbf{L}}}
\newcommand{\bD}{\ensuremath{\mathbf{D}}}
\newcommand{\bI}{\ensuremath{\mathbf{I}}}
\newcommand{\bx}{\ensuremath{\mathbf{x}}}
\newcommand{\bX}{\ensuremath{\mathbf{X}}}

\newcommand{\bZ}{\ensuremath{\mathbf{Z}}}

\usepackage[dvipsnames]{xcolor}

\title{Multimodal Clustering with Role Induced Constraints for Speaker Diarization}
\name{Nikolaos Flemotomos, Shrikanth Narayanan}
\address{
  Signal Analysis and Interpretation Laboratory\\
  University of Southern California, Los Angeles, CA, USA}
\email{flemotom@usc.edu, shri@ee.usc.edu}

\begin{document}
\maketitle
\begin{abstract}
Speaker clustering is an essential step in conventional speaker diarization systems and is typically addressed as an audio-only speech processing task. The language used by the participants in a conversation, however, carries additional information that can help improve the clustering performance. This is especially true in  conversational interactions, such as business meetings, interviews, and lectures, where specific roles assumed by interlocutors (manager, client, teacher, etc.) are often associated with distinguishable linguistic patterns. In this paper we propose to employ a supervised text-based model to extract speaker roles and then use this information to guide an audio-based spectral clustering step by imposing must-link and cannot-link constraints between segments. The proposed method is applied on two different domains, namely on medical interactions and on podcast episodes, and is shown to yield improved results when compared to the audio-only approach.
\end{abstract}

\vspace*{-2mm}
\section{Introduction}\vspace*{-.5mm}
Speaker diarization is the task of segmenting a multi-party speech signal into speaker-homogeneous regions and tagging them with speaker-specific labels~\cite{anguera2012review,park2022review}. Even though recently introduced end-to-end neural diarization offers simplicity and achieves remarkable results in some scenarios~\cite{fujita2019end,horiguchi2020end}, modular, clustering-based diarization is still widely used and has been an indispensable part of award-winning systems~\cite{medennikov2020stc,wang2021ustc}. In the conventional diarization approach, the speech signal is first segmented into chunks assumed to be speaker-homogeneous, in the sense that a single speaker is active therein. Fixed-dimensional speaker representations are then estimated for all the segments and their pairwise similarities are computed. Finally, a clustering algorithm gives the desired labeled speech segments. 

Even though it is generally assumed that no information is known a priori about the speakers, in practice we often need to deploy diarization systems in specific applications, and domain-dependent processing can be used to further improve the final performance. To that end, both the acoustic~\cite{wang2021ustc} and the linguistic~\cite{flemotomos20_odyssey_linguistically} streams of information have been successfully used to either adapt the models or modify the diarization pipeline. The language-based approach, where the transcripts of a recording are taken into consideration during diarization, is especially promising for interactions where speakers play dissimilar \textit{roles}.

In real-life conversations, interlocutors often assume specific roles which are associated with pre-defined objectives within a group~\cite{hare1994types}. Those in turn are linked to specific communication patterns which can be manifest through the language used within the conversations. For example, during an interview the \textit{interviewer} is expected to use more interrogative words than the \textit{interviewee} and during a flight the \textit{air traffic controller} is expected to give instructions to the \textit{pilot}, who responds by reading back the instructions in a particular manner. It should be noted that several role-playing conversations, such as interviews, clinical interactions, and court hearings, have been included in the evaluation data of recent diarization challenges~\cite{ryant2021third}.

This role-specific linguistic variability can be used to improve the performance of core speech processing tasks, including diarization. In our previous work, we used roles to estimate the profiles of the participants in a conversation and reduce the clustering problem into a classification one~\cite{flemotomos20_odyssey_linguistically}. In a related work, we ran audio-based speaker clustering and language-based role recognition in parallel and then combined the outputs through a meta-classifier~\cite{flemotomos2018combined}. However, those systems assume that each speaker is linked to a unique role during a conversation. Even though this is often a reasonable assumption (e.g., dialogues between a \textit{clinician} and a \textit{patient}), the systems cannot be easily generalized when a single speaker assumes multiple roles or when multiple speakers play the same role (e.g., trials with a single \textit{judge} and multiple \textit{prosecution witnesses}).

To overcome this limitation, we propose to exploit the linguistically extracted role information to impose constraints during audio-based clustering. Depending on the domain, we can impose must-link and/or cannot-link constraints, without the need for one-to-one correspondence between speakers and roles. Using manually-derived speaker-homogeneous segments with oracle transcriptions, we evaluate the effectiveness of the approach on the clustering performance by running experiments on two domains: i) dyadic clinical interactions, and ii) multi-party interactions from a weekly radio show.

\vspace*{-2mm}
\section{Background}\vspace*{-.7mm}
\subsection{Spectral clustering for speaker diarization}\label{subsec:spectral}\vspace*{-.8mm}
Clustering is one of the main components in modular speaker diarization. During that step, speech segments are grouped into same-speaker classes, usually following either a hierarchical agglomerative clustering (HAC)~\cite{sell18_interspeech}, or a spectral clustering~\cite{wang2018speaker} approach. This grouping is based on the pairwise similarities (e.g., cosine-based) between the $N$ segments to be clustered, which are stored in an affinity matrix $\hat\bW\in\mathbb{R}^{N\times N}$.

Having constructed the refined affinity matrix $\mathbf{W}$ (where refinements are explained later), spectral clustering is a technique that exploits the eigen-decomposition of $\bW$ to project the $N$ elements onto a suitable lower-dimensional space~\cite{ng2002spectral}. To do so, we define the degrees \mbox{$d_i\triangleq \sum_j{\bW_{ij}}$} and we construct the normalized Laplacian matrix \mbox{$\bL = \bI - \bD^{-1/2}\bW\bD^{-1/2}$}, where \mbox{$\bD=\text{diag}\{d_1,d_2,\cdots,d_N\}$}. Assuming we know the number of speakers $k$, we find the $k$ eigenvectors $\bx_1,\bx_2,\cdots,\bx_k$ corresponding to the $k$ smallest eigenvalues of $\bL$ and form the matrix $\bX=[\bx_1|\bx_2|\cdots|\bx_k]$. After normalizing the rows of $\bX$ to unit norm, we cluster them through a $k$-means algorithm and assign the $l$-th segment to speaker $s$ iff the $l$-th row is assigned to $s$.

In order to effectively use spectral clustering in diarization settings, several refinement operations have been proposed~\cite{wang2018speaker,park2019auto}, the most notable being \mbox{$p$-thresholding}. During that step, the $(100-p)\%$ largest values in each row of $\hat\bW$ are set to 1 and the rest are multiplied by a small constant~$\tau$ (soft thresholding), giving the modified matrix $\hat\bW_p$. Since this operation may break the symmetry property of the affinity matrix, we re-symmetrize it to get $\bW~=~\frac{1}{2}\left( \hat\bW_p + \hat\bW^T_p\right)$.

Instead of fixing a specific value $p$, an auto-tuning approach which uses the maximum eigengap of the Laplacian matrix can be followed~\cite{park2019auto}. The eigengap criterion has its roots in graph theory and is also used to estimate the number of clusters (speakers) $\hat k$, when this is not known a priori. $\bL$ is a positive semi-definite matrix with $N$ non-negative real eigenvalues $0=\lambda_1 \leq \lambda_2\leq \cdots\leq \lambda_N$. If $\bW$ is viewed as an adjacency matrix of a graph with $\hat k$ perfectly connected components, then $\hat k$ equals the multiplicity of the eigenvalue $\lambda_1=0$. In practical applications, where we do not expect perfect components, $\hat k$ is estimated by the maximum eigengap as $\hat k = \arg\max_k \frac{\lambda_{k+1}}{\lambda_{k}}$.

\vspace*{-1mm}
\subsection{Constrained spectral clustering}\label{subsec:constr_clustering}
Constrained clustering extends the traditional unsupervised learning paradigm of clustering by integrating supplemental information in the form of constraints~\cite{ganccarski2020constrained}. Even though several types of constraints have been explored, the most common ones are the instance-level relations, and in particular the must-link (ML) and cannot-link (CL) constraints. Under that viewpoint, if an ML (CL) constraint is imposed between two segments, then those segments must (must not) be in the same cluster. In speaker diarization, constrained clustering has been applied with constraints imposed either by human input~\cite{yu2017active}, or by acquired knowledge within a particular framework~\cite{bost2014constrained,kinoshita21_interspeech,tripathi2022turn2diarize}.

Constraints can be combined with several clustering algorithms, such as k-means~\cite{kinoshita2021integrating} or HAC~\cite{prokopalo2021active}. In this work we use a constrained spectral clustering approach, where constraints are integrated via the exhaustive and efficient constraint propagation (E$^2$CP) algorithm~\cite{lu2013exhaustive}, which was recently applied in diarization settings~\cite{tripathi2022turn2diarize}. Applying E$^2$CP, we can propagate an initial set of pairwise constraints to the entire session. In order to do so, we define a constraint matrix $\bZ\in\mathbb{R}^{N\times N}$, such that
\begin{equation}\label{eq:z_constr}
\bZ_{ij} = \left\lbrace\begin{array}{rl}
+1, & \text{if } \exists \text{ ML constraint between } i \text{ and } j  \\
-1, & \text{if } \exists \text{\hspace*{.4mm} CL\hspace*{.4mm} constraint between } i \text{ and } j \\
0,  & \text{if } \nexists \text{\hspace*{.4mm} any constraint between } i \text{ and } j
\end{array}  \right.    
\end{equation}

The elements of the affinity matrix $\hat\bW$ are then updated as 
\begin{equation}\label{eq:w_update}
\hat\bW_{ij} \leftarrow \left\lbrace\begin{array}{ll}
1 - (1-\bZ^*_{ij})(1-\hat\bW_{ij}), & \text{if } \bZ^*_{ij}\geq0 \\
(1+\bZ^*_{ij})\hat\bW_{ij} , & \text{if } \bZ^*_{ij}<0
\end{array}  \right.
\end{equation}
where $\bZ^*$ contains the constraints propagated to the entire session based on the initial set of constraints and is estimated as 
\begin{equation}\label{eq:alpha_constr}
\bZ^* = (1-\alpha)^2(\bI - \alpha \bar\bL)^{-1}\bZ(\bI - \alpha \bar\bL)^{-1}.
\end{equation}
$\bar\bL$ equals $\bar\bD^{-1/2}\hat\bW\bar\bD^{-1/2}$, where $\bar\bD$ is a diagonal matrix defined like $\bD$ in Section~\ref{subsec:spectral}, but using the degrees of $\hat\bW$. The constant $\alpha\in[0,1]$ is a tunable hyperparameter: a small value penalizes large changes between the initial pairwise constraints in $\bZ$ and the new constraints created during propagation, while a large value penalizes large changes between the neighboring segments in the graph described by $\hat\bW$. Note that for $\alpha=0$ we get $\bZ^*=\bZ$ which means we only rely on the initial constraints, and for $\alpha=1$ we get $\bZ^*=\mathbf{0}$, which means we completely ignore any constraint information. The constraint propagation and integration described here takes place before the refinement and spectral operations described in Section~\ref{subsec:spectral}. 

\vspace*{-2mm}
\section{Proposed Method}\label{sec:method}
\begin{figure}[ht]
    \centering
    \includegraphics[width=.9\columnwidth]{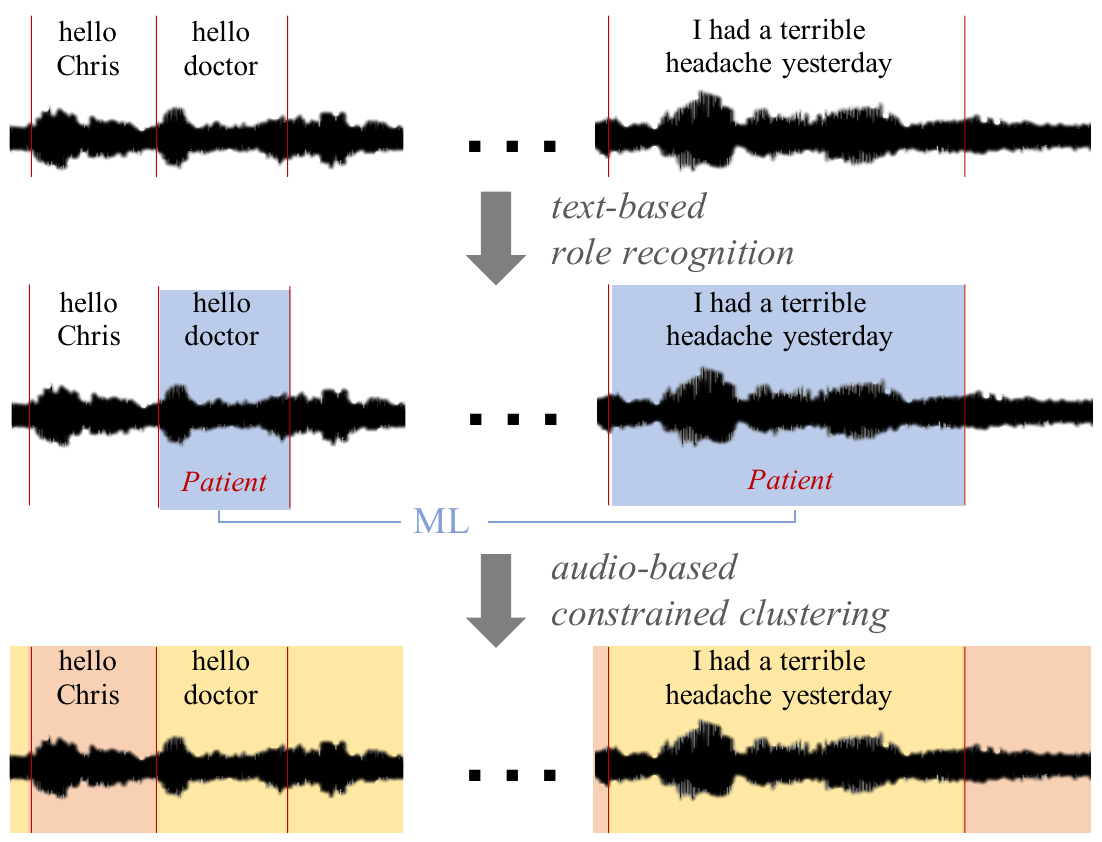}
    \caption{Two-step speaker clustering for role-playing interactions. Here, an ML constraint is imposed for two segments both associated with the role ``patient". Those segments have to be in the same cluster after the clustering step.\vspace*{-2mm}}   
    \label{fig:approach}
\end{figure}

In this work we propose to use a 2-step clustering for conversations where speakers assume distinct roles (Figure~\ref{fig:approach}).\vspace*{-2mm}

First, speaker roles are identified from text for each speech segment. To that end, we employ a BERT-based classifier~\cite{devlin2019bert}, where we add dropout and a softmax inference layer on top of a pre-trained BERT model and we fine-tune it for the task with in-domain data. If, after classification, we have complete role information available (i.e., each segment is associated with a distinct speaker role), we can directly get a purely text-based diarization result. However, there are multiple scenarios where only partial role information is available (e.g., we have sufficient data to only train a binary classifier to identify \textit{news anchor} vs. \textit{guest} in a broadcast news program with multiple potential \textit{guests} within a show). Additionally, we expect that there will be several segments where the linguistic content is not sufficient to robustly infer the associated speaker role.

So, we only use role information to impose suitable constraints for the following step of audio-based clustering, taking into account only segments where roles are identified with sufficient confidence. Even though neural classifiers tend to be over-confident about their decisions and softmax values are usually not a robust proxy of confidence scores, in practice we saw that we can use a softmax threshold as a threshold of confidence, as discussed in Section~\ref{sec:results}. For those segments where the confidence of their associated role is beyond some specified threshold, we impose ML and CL constraints, according to the domain we are working on. For instance, we can distinguish between the following general scenarios:

\begin{enumerate}
    \item \textit{different roles are always played by different speakers} (e.g., teacher vs. students): apply a CL constraint between segments associated with different speaker roles, 
    \item \textit{different speakers always play different roles} (e.g., anchor vs. interviewer vs. guest, where anchor and interviewer might be the same person): apply an ML constraint between segments associated with the same role, 
    \item \textit{one-to-one correspondence between speakers and roles} (e.g., doctor vs. patient): apply both CL and ML constraints as in cases~(1) and~(2).
\end{enumerate}

Different types of domain-specific strategies can also be followed. The constraints are then integrated in a spectral clustering algorithm as described in Sections~\ref{subsec:spectral} and~\ref{subsec:constr_clustering}.

\vspace*{-2mm}
\section{Datasets}
We evaluate the proposed speaker clustering approach on two domains with role-playing interactions, as detailed below.

\vspace*{1mm}
{\bf Psychotherapy sessions.} We use a collection of psychotherapy sessions recorded at a university counseling center (UCC), and specifically the sessions in the sets denoted as UCC$_{train}$, UCC$_{dev}$ and UCC$_{test_1}$ in~\cite{flemotomos2021automated}. Each session is a dyadic conversation between a \textit{therapist} and a \textit{patient}, thus falls under case (3) according to the categorization of Section~\ref{sec:method}. The dataset comprises 97 participants (23 therapists and 74 patients), with no speaker overlap between the train/dev/eval sets. The sessions have been professionally transcribed and the transcribed segments have been forced-aligned with the audio. 
More details on the dataset are provided in Table~\ref{table:ucc_data}. \vspace*{-1.5mm}

\begin{table}[ht]
\caption{Size of the UCC dataset.\vspace*{-3mm}}
\label{table:ucc_data}
\centering
    \begin{tabular}{llll}
        \toprule
         & \textbf{train} & \textbf{dev} & \textbf{eval} \\
        \midrule
        \#sessions & 50 & 26 & 20 \\
        \midrule
        \#segments - therapist & 8,766 & 3,959 & 4,146 \\
        \#segments - patient & 9,052 & 4,246 & 4,245 \\
        \midrule
        segment duration (mean) & 7.8sec & 8.7sec &6.4sec\\
        \#words per segment (mean) & 21.4 & 22.3 & 18.8 \\
        \bottomrule
    \end{tabular}
\end{table}\vspace*{-2mm}

{\bf Podcast episodes. }\textit{This American Life}\footnote{\url{www.thisamericanlife.org}} (TAL) is a weekly podcast and radio show. 
The authors in~\cite{mao20b_interspeech} have curated a dataset of 663 TAL episodes aired between 1995 and 2020. 
We use the audio-aligned utterances provided, with the recommended train/dev/eval split, and with the archived audio standardized as in~\cite{mao20b_interspeech}. Each episode features on average 17.7 speakers (std=8.7) with variable speaking times. The dataset, described in Table~\ref{table:tal_data}, has been annotated with speaker IDs and with three speaker roles, those of \textit{host}, \textit{interviewer}, and \textit{subject}. However, the provided role information was not helpful for our purposes, since multiple speakers may play the same role within an episode and, at the same time, a single speaker may play multiple roles (with some episodes having the same speaker occasionally playing all 3 roles). So, we chose to annotate as \textit{host} utterances only the ones spoken by Ira Glass and assign all the other utterances to a \textit{non-host} speaker role. Ira Glass is the host and executive producer of the show and speaks for $18.6\%$ of the time in the entire dataset\footnote{For reference, the second single most-talking speaker of the dataset is Nancy Updike, speaking for $1.6\%$ of the time.}. Since different roles always denote different speakers (but the inverse does not hold since there are multiple \textit{non-host} speakers), 
this dataset falls under case (1) according to the categorization given in Section~\ref{sec:method}.
Of course, we should note that, since this annotation strategy is speaker-dependent, the role recognition algorithm applied is also expected to capture speaker-specific, and not purely role-specific, information.
\vspace*{-1.5mm} 

\begin{table}[ht]
\caption{Size of the TAL dataset.\vspace*{-3mm}}
\label{table:tal_data}
\centering
    \begin{tabular}{llll}
        \toprule
         & \textbf{train} & \textbf{dev} & \textbf{eval} \\
        \midrule
        \#episodes & 593 & 34 & 36 \\
        \midrule
        \#segments - host & 26,523 & 1,765 & 1,317 \\
        \#segments - non-host & 119,295 & 6,869 & 8,039 \\
        \midrule
        segment duration (mean) & 14.1sec & 13.7sec & 13.4sec\\
        \#words per segment (mean) & 37.7 &  36.6&  36.6\\
        \bottomrule
    \end{tabular}
\end{table}\vspace*{-2mm}

\section{Experiments and Results}\label{sec:results}
\subsection{Experimental setup}\vspace*{-.5mm}
We run experiments using manually derived speaker segments and transcriptions to avoid propagating potential errors from automated segmentation and speech recognition modules. We standardize the text by stripping punctuation, removing non-verbal vocalizations and converting all letters to lower case. 

We build the binary role classifiers (\textit{therapist} vs. \textit{patient} and \textit{host} vs. \textit{non-host}) using TensorFlow~\cite{abadi2016tensorflow} with the pre-trained uncased English BERT-base model provided in TensorFlow model garden~\cite{tensorflowmodelgarden2020}. Since very short segments are not expected to have sufficient role-related information, during fine-tuning we only take into account segments containing at least 5 words ($65.58\%$ of the available training segments for UCC and $88.65\%$ of the available training segments for TAL). We fine-tune the models for 2 epochs, using the Adam optimizer with decoupled weight decay~\cite{loshchilov2019decoupled}, with initial learning rate equal to $2\cdot10^{-5}$ and with a warm-up stage lasting for the first $10\%$ of the training time. The mini-batch size is set to 16 segments and the maximum allowed segment length is set to 128 tokens\footnote{according to the default WordPiece-based BERT tokenizer}, which means that $2.91\%$ of the initial training UCC segments and $2.06\%$ of the initial training TAL segments are cropped.

The speaker representation of the segments is based on the widely used x-vectors~\cite{snyder2018xvector}. To that end, the pre-trained VoxCeleb x-vector extractor from Kaldi~\cite{povey2011kaldi} is used\footnote{\url{kaldi-asr.org/models/m7}} and a single normalized x-vector is extracted per segment. The segments are clustered following the described constrained spectral clustering approach\footnote{\url{github.com/wq2012/SpectralCluster}} with cosine-based affinities and with ML and/or CL constraints imposed according to the predicted associated roles. For the UCC dataset, which features dyadic interactions, we group all the segments into two clusters, while for TAL we estimate the number of speakers using the eigengap criterion searching in the range 2--50. The value of $p$ for $p$-thresholding is found through auto-tuning~\cite{park2019auto}, searching in the range 40--95, and we use soft thresholding with $\tau=0.01$~\cite{wang2018speaker}. 

All the results are reported on the eval subsets of the data. Diarization is evaluated with respect to the diarization error rate (DER), estimated with the pyannote.metrics library~\cite{pyannote.metrics} without allowing any tolerance collar around segment boundaries. DER incorporates three sources of error; false alarms, 
missed speech, 
and speaker confusion. 
However, segmentation is always the oracle one provided by human annotators and, since there is almost no speaker overlap in our datasets, by DER we essentially estimate speaker confusion (false alarm is always $0$ and missed speech is $0.02\%$ for UCC and $0.13\%$ for TAL). \vspace*{-1mm}

\subsection{Results and discussion}
\begin{figure}[ht]
\centering
\includegraphics[width=.9\columnwidth]{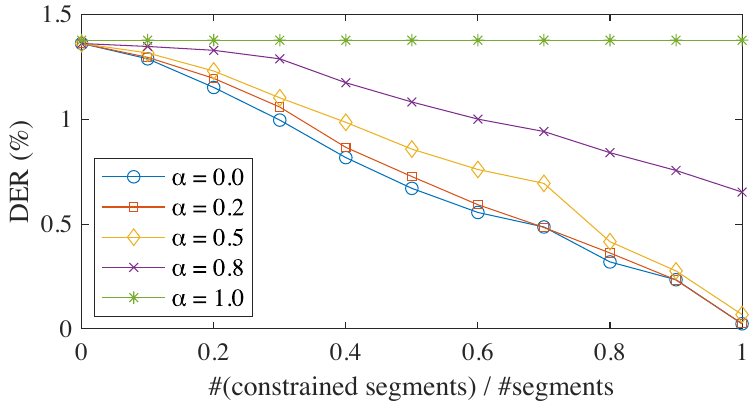}\vspace*{-2mm}
\caption{Diarization error rate -- lower is better -- for the UCC dataset as a function of the (normalized) number of constraints, always providing oracle role information to build the constraints, for different values of $\alpha$ in eq.~\eqref{eq:alpha_constr}.\vspace*{-5mm}}
\label{fig:oracle_constraints}
\end{figure}

\vspace*{-.5mm}
If we have perfect role information for all the segments and if there is one-to-one correspondence between roles and speakers (e.g., one \textit{therapist} vs. one \textit{patient} in UCC), we can get a perfect diarization result in terms of speaker confusion. In the framework of constrained spectral clustering, this can be done by filling $\bZ$ in eq.~\eqref{eq:z_constr} with all the corresponding constraints and setting $\alpha=0$ in eq.~\eqref{eq:alpha_constr} so that $\bZ^*=\bZ$. This is reflected in Figure~\ref{fig:oracle_constraints} where we see how DER changes as we provide more oracle constraints to the algorithm. We note that this is similar to the expected behavior when constraints are added in the form of human supervision.

Without having access to the oracle role information, we have to rely on a segment-level role classifier. After fine-tuning, the classification accuracy of the BERT-based classifiers, when only segments with at least 5 words are taken into consideration, is $83.22\%$ for UCC (vs. $53.50\%$ for a majority-class baseline) and $90.92\%$ for TAL (vs. $85.41\%$ for a majority-class baseline). Even though the classifiers provide reasonable results, we need to ensure that constraints are imposed only on segments which are confidently linked to some role. For this work, apart from only using segments longer than a specified duration (here, containing at least 5 words) to ensure some minimal linguistic content, we use the softmax values associated with the predicted roles as a proxy of the confidence level.

\begin{figure}[ht]
\centering
\begin{subfigure}{.49\columnwidth}
\includegraphics[width=\textwidth]{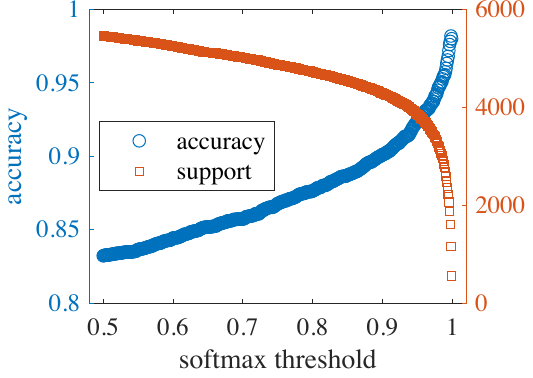}
\caption{UCC\vspace*{-2.2mm}}
\end{subfigure}
\begin{subfigure}{.49\columnwidth}
\includegraphics[width=\textwidth]{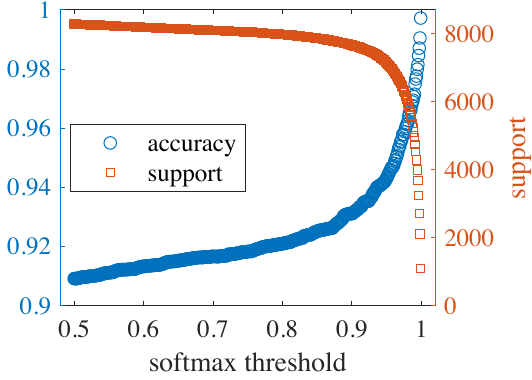}
\caption{TAL\vspace*{-2mm}}
\end{subfigure}
\caption{Classification accuracy and support for the BERT-based classifiers when only segments with associated softmax value above some threshold are considered.\vspace*{-2mm}}
\label{fig:softmax_thres}
\end{figure}

As shown in Figure~\ref{fig:softmax_thres}, the softmax value can indeed act as a reasonable proxy of confidence for our purposes. In particular, if we only consider segments where the corresponding softmax value is above some threshold, accuracy increases monotonically as a function of the threshold. However, the choice of the threshold value is a trade-off decision between accuracy and adequate support so that we have a sufficient number of constraints. For this work, we choose a threshold equal to 0.980 for UCC (accuracy = 94.66$\%$, support = 3,222) and equal to 0.995 for TAL (accuracy = 98.15$\%$, support = 3,674), imposing constraints on around $40\%$ of the segments in both cases.\vspace*{-.4mm}

After constructing the constraint matrix based on the described role classification only for the segments with sufficient role classification confidence, we perform the constrained spectral clustering algorithm. 
The results\footnote{Those results are for $\alpha=0.75$ for UCC and $\alpha=0.50$ for TAL.} are reported in the second column of Table~\ref{table:diar_res}. For comparison, we also provide results for the two extreme cases: i) following a conventional, unconstrained spectral clustering, which ignores any language-based information and ii) following a language-only classification, by using the results of the BERT-based classifier for all the segments, which ignores any audio-based information. 

\begin{table}[ht]
\caption{Diarization error rate ($\%$) -- lower is better -- using unconstrained audio-only clustering, constrained clustering with role-induced constraints, and language-only classification.}
\label{table:diar_res}
\centering
    \begin{tabular}{cccc}
        \toprule
        & \spcell{\textbf{unconstrained}\\\textbf{clustering}\\(audio-only)} & \spcell{\textbf{constrained}\\\textbf{clustering}\\(multimodal)} & \spcell{\textbf{role-based}\\\textbf{classification}\\(language-only)}\\
        \midrule
        UCC & 1.38 & 1.31 & 10.34\hspace*{.2cm} \\
        TAL & 42.22 & 23.86 & 63.01$^*$ \\
        \bottomrule
    \end{tabular}\\
    {\raggedright $^*$results contain 2 speakers (due to the binary classification) \par}
\end{table}\vspace*{-7mm}

In the case of the UCC data, our approach yields a small improvement (5.1$\%$ relative) compared to the unconstrained baseline. We additionally found that adding more constraints (selecting a smaller softmax threshold as our confidence criterion) leads to worse performance. Comparing this finding to the results displayed in Figure~\ref{fig:oracle_constraints} with oracle constraints, where error approaches 0 given a large number of constraints, we realize that our method is sensitive to the performance of the role classifier. This is because any classification errors can be easily propagated to the clustering step (Figure~\ref{fig:approach}). This becomes more evident when we constrain all the segments, relying only on the linguistic stream of information (last column of Table~\ref{table:diar_res}).     

For the TAL data we observe a substantial improvement going from unconstrained to constrained clustering, suggesting that in scenarios with a large number of speakers, even partial role-based information 
can provide useful cues that robustly guide the subsequent clustering. In more detail, we found that the imposed constraints changed the final Laplacian matrices in a way that led to the detection of more speakers per episode. The severe performance degradation with the language-only approach is expected, since the results in that case only contain two speakers (since we only have two role classes), even though each TAL episode features multiple participants. \vspace*{-1.5mm}

\section{Conclusion}\vspace*{-.5mm}
We proposed to integrate text-based constraints within audio-based clustering to improve speaker diarization in role-playing conversational interactions. We implemented a BERT-based role classifier relying on text data and used its output to construct a constraint matrix for use within constrained spectral clustering. Experimental results showed that, after applying a softmax-based confidence criterion, performance can be improved both in cases of one-to-one correspondence between speakers and roles and when only partial role information is available, thus overcoming limitations of previous works.

We performed all our experiments using oracle textual information and oracle speaker segmentation. As part of future work, we intend to study the final performance with respect to the word error rate of a speech recognizer, and to errors due to non-ideal segmentation. Even though diarization is widely viewed as a pre-processing step before an automatic speech recognition (ASR) system, this conventional paradigm is often challenged by modern pipelines where the diarization algorithm utilizes ASR-derived information (e.g.,~\cite{park19_interspeech,yoshioka19_interspeech}). Speaker segmentation could be included as a separate module (e.g.,~\cite{flemotomos20_odyssey_linguistically}), or incorporated within the role recognizer in a named entity recognition (NER)-like approach (e.g.,~\cite{zuluaga2021bertraffic}). Future work can also investigate different types of role-induced constraints. Even though we focused on linguistic characteristics, role-specific behaviors can also be manifest through acoustic, structural, or visual cues, all of which can be potentially used within the framework of role-dependent constrained speaker clustering.

\vfill

\pagebreak
\bibliographystyle{IEEEtran}
\bibliography{references}

\begin{thebibliography}{10}
\providecommand{\url}[1]{#1}
\csname url@samestyle\endcsname
\providecommand{\newblock}{\relax}
\providecommand{\bibinfo}[2]{#2}
\providecommand{\BIBentrySTDinterwordspacing}{\spaceskip=0pt\relax}
\providecommand{\BIBentryALTinterwordstretchfactor}{4}
\providecommand{\BIBentryALTinterwordspacing}{\spaceskip=\fontdimen2\font plus
\BIBentryALTinterwordstretchfactor\fontdimen3\font minus
  \fontdimen4\font\relax}
\providecommand{\BIBforeignlanguage}[2]{{%
\expandafter\ifx\csname l@#1\endcsname\relax
\typeout{** WARNING: IEEEtran.bst: No hyphenation pattern has been}%
\typeout{** loaded for the language `#1'. Using the pattern for}%
\typeout{** the default language instead.}%
\else
\language=\csname l@#1\endcsname
\fi
#2}}
\providecommand{\BIBdecl}{\relax}
\BIBdecl

\bibitem{anguera2012review}
X.~Anguera, S.~Bozonnet, N.~Evans, C.~Fredouille, G.~Friedland, and O.~Vinyals,
  ``Speaker diarization: A review of recent research,'' \emph{IEEE Transactions
  on Audio, Speech, and Language Processing}, vol.~20, no.~2, pp. 356--370,
  2012.

\bibitem{park2022review}
T.~J. Park, N.~Kanda, D.~Dimitriadis, K.~J. Han, S.~Watanabe, and S.~Narayanan,
  ``A review of speaker diarization: Recent advances with deep learning,''
  \emph{Computer Speech \& Language}, vol.~72, p. 101317, 2022.

\bibitem{fujita2019end}
Y.~Fujita, N.~Kanda, S.~Horiguchi, K.~Nagamatsu, and S.~Watanabe, ``End-to-end
  neural speaker diarization with permutation-free objectives,'' in \emph{Proc.
  Interspeech}, 2019, pp. 4300--4304.

\bibitem{horiguchi2020end}
S.~Horiguchi, Y.~Fujita, S.~Watanabe, Y.~Xue, and K.~Nagamatsu, ``End-to-end
  speaker diarization for an unknown number of speakers with encoder-decoder
  based attractors,'' in \emph{Proc. Interspeech}, 2020, pp. 269--273.

\bibitem{medennikov2020stc}
I.~Medennikov, M.~Korenevsky, T.~Prisyach, Y.~Khokhlov, M.~Korenevskaya,
  I.~Sorokin, T.~Timofeeva, A.~Mitrofanov, A.~Andrusenko, I.~Podluzhny
  \emph{et~al.}, ``The {STC} system for the {CHiME-6} challenge,'' in
  \emph{Proc. International Workshop on Speech Processing in Everyday
  Environments (CHiME)}, 2020, pp. 36--41.

\bibitem{wang2021ustc}
Y.~Wang, M.~He, S.~Niu, L.~Sun, T.~Gao, X.~Fang, J.~Pan, J.~Du, and C.-H. Lee,
  ``{USTC-NELSLIP} system description for {DIHARD-III} challenge,'' \emph{arXiv
  preprint arXiv:2103.10661}, 2021.

\bibitem{flemotomos20_odyssey_linguistically}
N.~Flemotomos, P.~Georgiou, and S.~Narayanan, ``Linguistically aided speaker
  diarization using speaker role information,'' in \emph{Proc. The Speaker and
  Language Recognition Workshop (Odyssey)}, 2020, pp. 117--124.

\bibitem{hare1994types}
A.~P. Hare, ``Types of roles in small groups: A bit of history and a current
  perspective,'' \emph{Small Group Research}, vol.~25, no.~3, pp. 433--448,
  1994.

\bibitem{ryant2021third}
N.~Ryant, P.~Singh, V.~Krishnamohan, R.~Varma, K.~Church, C.~Cieri, J.~Du,
  S.~Ganapathy, and M.~Liberman, ``The third {DIHARD} diarization challenge,''
  \emph{arXiv preprint arXiv:2012.01477}, 2021.

\bibitem{flemotomos2018combined}
N.~Flemotomos, P.~Papadopoulos, J.~Gibson, and S.~Narayanan, ``Combined speaker
  clustering and role recognition in conversational speech,'' in \emph{Proc.
  Interspeech}, 2018, pp. 1378--1382.

\bibitem{sell18_interspeech}
G.~Sell, D.~Snyder, A.~McCree, D.~Garcia-Romero, J.~Villalba, M.~Maciejewski,
  V.~Manohar, N.~Dehak, D.~Povey, S.~Watanabe, and S.~Khudanpur, ``Diarization
  is hard: Some experiences and lessons learned for the {JHU} team in the
  inaugural {DIHARD} challenge,'' in \emph{Proc. Interspeech}, 2018, pp.
  2808--2812.

\bibitem{wang2018speaker}
Q.~Wang, C.~Downey, L.~Wan, P.~A. Mansfield, and I.~L. Moreno, ``Speaker
  diarization with {LSTM},'' in \emph{Proc. IEEE International Conference on
  Acoustics, Speech and Signal Processing (ICASSP)}, 2018, pp. 5239--5243.

\bibitem{ng2002spectral}
A.~Y. Ng, M.~I. Jordan, and Y.~Weiss, ``On spectral clustering: Analysis and an
  algorithm,'' in \emph{Advances in neural information processing systems},
  2002, pp. 849--856.

\bibitem{park2019auto}
T.~J. Park, K.~J. Han, M.~Kumar, and S.~Narayanan, ``Auto-tuning spectral
  clustering for speaker diarization using normalized maximum eigengap,''
  \emph{IEEE Signal Processing Letters}, vol.~27, pp. 381--385, 2019.

\bibitem{ganccarski2020constrained}
P.~Gan{\c{c}}arski, T.-B.-H. Dao, B.~Cr{\'e}milleux, G.~Forestier, and
  T.~Lampert, ``Constrained clustering: Current and new trends,'' in \emph{A
  guided tour of artificial intelligence research}.\hskip 1em plus 0.5em minus
  0.4em\relax Springer, 2020, pp. 447--484.

\bibitem{yu2017active}
C.~Yu and J.~H. Hansen, ``Active learning based constrained clustering for
  speaker diarization,'' \emph{IEEE/ACM Transactions on Audio, Speech, and
  Language Processing}, vol.~25, no.~11, pp. 2188--2198, 2017.

\bibitem{bost2014constrained}
X.~Bost and G.~Linares, ``Constrained speaker diarization of {TV} series based
  on visual patterns,'' in \emph{IEEE Spoken Language Technology Workshop
  (SLT)}, 2014, pp. 390--395.

\bibitem{kinoshita21_interspeech}
K.~Kinoshita, M.~Delcroix, and N.~Tawara, ``Advances in integration of
  end-to-end neural and clustering-based diarization for real conversational
  speech,'' in \emph{Proc. Interspeech}, 2021, pp. 3565--3569.

\bibitem{tripathi2022turn2diarize}
A.~Tripathi, H.~Lu, H.~Sak, I.~L. Moreno, Q.~Wang, and W.~Xia,
  ``Turn-to-diarize: Online speaker diarization constrained by transformer
  transducer speaker turn detection,'' in \emph{Proc. IEEE International
  Conference on Acoustics, Speech and Signal Processing (ICASSP) -- to appear},
  2022.

\bibitem{kinoshita2021integrating}
K.~Kinoshita, M.~Delcroix, and N.~Tawara, ``Integrating end-to-end neural and
  clustering-based diarization: Getting the best of both worlds,'' in
  \emph{IEEE International Conference on Acoustics, Speech and Signal
  Processing (ICASSP)}, 2021, pp. 7198--7202.

\bibitem{prokopalo2021active}
Y.~Prokopalo, M.~Shamsi, L.~Barrault, S.~Meignier, and A.~Larcher, ``Active
  correction for speaker diarization with human in the loop,'' in \emph{Proc.
  Iberspeech}, 2021.

\bibitem{lu2013exhaustive}
Z.~Lu and Y.~Peng, ``Exhaustive and efficient constraint propagation: A
  graph-based learning approach and its applications,'' \emph{International
  journal of computer vision}, vol. 103, no.~3, pp. 306--325, 2013.

\bibitem{devlin2019bert}
J.~Devlin, M.-W. Chang, K.~Lee, and K.~Toutanova, ``{BERT}: Pre-training of
  deep bidirectional transformers for language understanding,'' in \emph{Proc.
  Conference of the North American Chapter of the Association for Computational
  Linguistics: Human Language Technologies, Volume 1 (Long and Short Papers)},
  2019, pp. 4171--4186.

\bibitem{flemotomos2021automated}
N.~Flemotomos, V.~R. Martinez, Z.~Chen, K.~Singla, V.~Ardulov, R.~Peri, D.~D.
  Caperton, J.~Gibson, M.~J. Tanana, P.~Georgiou \emph{et~al.}, ``Automated
  evaluation of psychotherapy skills using speech and language technologies,''
  \emph{Behavior Research Methods}, 2021.

\bibitem{mao20b_interspeech}
H.~H. Mao, S.~Li, J.~McAuley, and G.~W. Cottrell, ``Speech recognition and
  multi-speaker diarization of long conversations,'' in \emph{Proc.
  Interspeech}, 2020, pp. 691--695.

\bibitem{abadi2016tensorflow}
M.~Abadi, P.~Barham, J.~Chen, Z.~Chen, A.~Davis, J.~Dean, M.~Devin,
  S.~Ghemawat, G.~Irving, M.~Isard \emph{et~al.}, ``{TensorFlow}: A system for
  large-scale machine learning,'' in \emph{USENIX symposium on operating
  systems design and implementation (OSDI)}, 2016, pp. 265--283.

\bibitem{tensorflowmodelgarden2020}
H.~Yu, C.~Chen, X.~Du, Y.~Li, A.~Rashwan, L.~Hou, P.~Jin, F.~Yang, F.~Liu,
  J.~Kim, and J.~Li, ``{TensorFlow Model Garden},''
  \url{https://github.com/tensorflow/models}, 2020.

\bibitem{loshchilov2019decoupled}
I.~Loshchilov and F.~Hutter, ``Decoupled weight decay regularization,'' in
  \emph{Proc. International Conference on Learning Representations (ICLR)},
  2019.

\bibitem{snyder2018xvector}
D.~Snyder, D.~Garcia-Romero, G.~Sell, D.~Povey, and S.~Khudanpur, ``X-vectors:
  Robust dnn embeddings for speaker recognition,'' in \emph{Proc. IEEE
  International Conference on Acoustics, Speech and Signal Processing
  (ICASSP)}, 2018, pp. 5329--5333.

\bibitem{povey2011kaldi}
D.~Povey, A.~Ghoshal, G.~Boulianne, L.~Burget, O.~Glembek, N.~Goel,
  M.~Hannemann, P.~Motlicek, Y.~Qian, P.~Schwarz, J.~Silovsky, G.~Stemmer, and
  K.~Vesely, ``The kaldi speech recognition toolkit,'' in \emph{Proc. Workshop
  on Automatic Speech Recognition and Understanding (ASRU)}, 2011.

\bibitem{pyannote.metrics}
H.~Bredin, ``{pyannote.metrics: a toolkit for reproducible evaluation,
  diagnostic, and error analysis of speaker diarization systems},'' in
  \emph{Proc. Interspeech}, 2017, pp. 3587--3591.

\bibitem{park19_interspeech}
T.~J. Park, K.~J. Han, J.~Huang, X.~He, B.~Zhou, P.~Georgiou, and S.~Narayanan,
  ``Speaker diarization with lexical information,'' in \emph{Proc.
  Interspeech}, 2019, pp. 391--395.

\bibitem{yoshioka19_interspeech}
T.~Yoshioka, D.~Dimitriadis, A.~Stolcke, W.~Hinthorn, Z.~Chen, M.~Zeng, and
  X.~Huang, ``Meeting transcription using asynchronous distant microphones,''
  in \emph{Proc. Interspeech 2019}, 2019, pp. 2968--2972.

\bibitem{zuluaga2021bertraffic}
J.~Zuluaga-Gomez, S.~S. Sarfjoo, A.~Prasad, I.~Nigmatulina, P.~Motlicek,
  O.~Ohneiser, and H.~Helmke, ``{BERTraffic}: A robust {BERT}-based approach
  for speaker change detection and role identification of air-traffic
  communications,'' \emph{arXiv preprint arXiv:2110.05781}, 2021.

\end{thebibliography}

\end{document}